\documentclass[journal]{IEEEtran}
\usepackage{url}
\usepackage[utf8]{inputenc}
\usepackage{xcolor}
\usepackage{amsmath}
\usepackage{amssymb}

\usepackage{tablefootnote}
\usepackage{booktabs}
\usepackage{tabularx}
\usepackage{tikz}
\usepackage{pgfplots}
\pgfplotsset{compat=newest} 
\pgfplotsset{plot coordinates/math parser=false} 
\newlength\fheight
\newlength\fwidth
\usetikzlibrary{plotmarks,patterns,decorations.pathreplacing,backgrounds,calc,arrows,arrows.meta,spy,matrix}
\usepgfplotslibrary{patchplots,groupplots}
\usepackage{tikzscale}
\usepackage{siunitx}

\usepackage{tikz}
    \usetikzlibrary{shapes.arrows}

\usepackage{pgfplots}
\usepackage{adjustbox}

\usepackage{gincltex}
\usepgfplotslibrary{fillbetween}
\usetikzlibrary{plotmarks}

\usepackage{multirow}
\usepackage[font=scriptsize]{subcaption}
\usepackage[font=footnotesize]{caption}

\usepackage{mathtools}

\usepackage{dblfloatfix}    
\usepackage{colortbl}

\usepackage{makecell}
\usepackage{diagbox}
\usepackage{tikz-qtree}
\usetikzlibrary{trees} 
\usepackage{cite}
\usepackage{bbold}
\usepackage{bbm}

\usepackage{array}
\newcolumntype{?}{!{\vrule width 1.5pt}}
\newcolumntype{P}[1]{>{\centering\arraybackslash}p{#1}}
\usepackage{makecell}
\usepackage{mdframed}
\usepackage[many]{tcolorbox}
\usepackage{enumitem}
\usepackage{booktabs}

\definecolor{violet}{rgb}{0.6,0,0.6}%
\definecolor{orange_D}{rgb}{1,0.3,0}%
\definecolor{cyan}{rgb}{0,0.67,0.64}%
\definecolor{red}{rgb}{0.9,0,0}%
\definecolor{green}{rgb}{0,0.8,0}%
\definecolor{yellow}{rgb}{1,0.8,0}

\newtcbox{\mybox}[1][]{nobeforeafter,math upper,tcbox raise base,
  enhanced,frame hidden,boxrule=0pt,interior style={top color=green!10!white,
  bottom color=green!10!white,middle color=green!50!yellow},
  fuzzy halo=1pt with green,drop large lifted shadow,#1}

\def \fwidth{0.8\columnwidth}
\def \fheight {0.4\columnwidth}

\usepackage{siunitx}
\usetikzlibrary{fadings}

\tikzfading[name=middle,
            top color=transparent!100,
            bottom color=transparent!100,
            middle color=transparent!20]

\usetikzlibrary{arrows,automata,calc,shapes, positioning,shadows,shadows.blur,shapes.geometric}

\usepackage[acronyms,nonumberlist,nopostdot,nomain,nogroupskip]{glossaries}

\newacronym{3gpp}{3GPP}{3rd Generation Partnership Project}
\newacronym{adc}{ADC}{Analog to Digital Converter}
\newacronym{5g}{5G}{5th generation}
\newacronym{6g}{6G}{6th generation}
\newacronym{aimd}{AIMD}{Additive Increase Multiplicative Decrease}
\newacronym{am}{AM}{Acknowledged Mode}
\newacronym{amc}{AMC}{Adaptive Modulation and Coding}
\newacronym{aqm}{AQM}{Active Queue Management}
\newacronym{awgn}{AGWN}{Additive White Gaussian Noise}
\newacronym{balia}{BALIA}{Balanced Link Adaptation}
\newacronym{bdp}{BDP}{Bandwidth-Delay Product}
\newacronym{bf}{BF}{beamforming}
\newacronym{cc}{CC}{Congestion Control}
\newacronym{cdf}{CDF}{Cumulative Distribution Function}
\newacronym{cn}{CN}{Core Network}
\newacronym{cqi}{CQI}{Channel Quality Information}
\newacronym{sctp}{SCTP}{Stream Control Transmission Protocol}
\newacronym{tls}{TLS}{Transport Layer Security}
\newacronym{bbr}{BBR}{bottleneck bandwidth and round-trip propagation time}
\newacronym{cp}{CP}{Control Plane}
\newacronym{csirs}{CSI-RS}{Channel State Information - Reference Signal}
\newacronym{dc}{DC}{Dual Connectivity}
\newacronym{fov}{FoV}{field of view}
\newacronym{rb}{RB}{Resource Block}
\newacronym{dce}{DCE}{Direct Code Execution}
\newacronym{dci}{DCI}{Downlink Control Information}
\newacronym{udp}{UDP}{User Datagram Protocol}
\newacronym{dl}{DL}{Downlink}
\newacronym{dmr}{DMR}{Deadline Miss Ratio}
\newacronym{dmrs}{DMRS}{DeModulation Reference Signal}
\newacronym{e2e}{E2E}{End-to-End}
\newacronym{ppp}{PPP}{Poission Point Process}
\newacronym{si}{SI}{Study Item}
\newacronym{ecn}{ECN}{Explicit Congestion Notification}
\newacronym{edf}{EDF}{Earliest Deadline First}
\newacronym{enb}{eNB}{eNodeB}
\newacronym{epc}{EPC}{Evolved Packet Core}
\newacronym{es}{ES}{Edge Server}
\newacronym{cav}{CAV}{Connected and Autonomous Vehicle}
\newacronym{fdma}{FDMA}{Frequency Division Multiple Access}
\newacronym{fdd}{FDD}{Frequency Division Duplexing}
\newacronym{upa}{UPA}{Uniform Planar Array}
\newacronym[firstplural=radio access technologies (RATs)]{rat}{RAT}{radio access technology}
\newacronym[firstplural=Radio Access Technology (RTs)]{rt}{RT}{Radio Technology}
\newacronym{fs}{FS}{Fast Switching}
\newacronym{isd}{ISD}{inter-site distance}
\newacronym{ftp}{FTP}{File Transfer Protocol}
\newacronym{gnb}{gNB}{Next Generation Node Base}
\newacronym{harq}{HARQ}{Hybrid Automatic Repeat reQuest}
\newacronym{hetnet}{HetNet}{Heterogeneous Network}
\newacronym{hh}{HH}{Hard Handover}
\newacronym{hol}{HOL}{Head-of-Line}
\newacronym{ia}{IA}{Initial Access}
\newacronym{imt}{IMT}{International Mobile Telecommunication}
\newacronym{iot}{IoT}{Internet of things}
\newacronym{jnd}{JND}{Just Noticeable Difference}
\newacronym{los}{LOS}{line of sight}
\newacronym{lte}{LTE}{Long Term Evolution}
\newacronym{m2m}{M2M}{Machine to Machine}
\newacronym{mac}{MAC}{Medium Access Control}
\newacronym{mc}{MC}{Multi-Connectivity}
\newacronym{mcs}{MCS}{modulation and coding scheme}
\newacronym{mec}{MEC}{Mobile Edge Cloud}
\newacronym{mi}{MI}{Mutual Information}
\newacronym{mimo}{MIMO}{Multiple Input Multiple Output}
\newacronym{mmwave}{mmWave}{millimeter wave}
\newacronym{mptcp}{MPTCP}{Multipath TCP}
\newacronym{mr}{MR}{Maximum Rate}
\newacronym{mss}{MSS}{Maximum Segment Size}
\newacronym{mtd}{MTD}{Machine-Type Device}
\newacronym{mtu}{MTU}{Maximum Transmission Unit}
\newacronym{nfv}{NFV}{Network Function Virtualization}
\newacronym{vnf}{VNF}{ Virtualization Network Function}
\newacronym{sdn}{SDN}{software defined networking}
\newacronym{nlos}{NLOS}{Non Line of Sight}
\newacronym{nlosb}{NLOSb}{Building Non Line of Sight}
\newacronym{nlosv}{NLOSv}{Vehicle Non Line of Sight}
\newacronym{nr}{NR}{New Radio}
\newacronym{ofdm}{OFDM}{Orthogonal Frequency Division Multiplexing}
\newacronym{pdcch}{PDCCH}{Physical Downlonk Control Channel}
\newacronym{pdcp}{PDCP}{Packet Data Convergence Protocol}
\newacronym{pdsch}{PDSCH}{Physical Downlink Shared Channel}
\newacronym{pdu}{PDU}{Packet Data Unit}
\newacronym{pf}{PF}{Proportional Fair}
\newacronym{pgw}{PGW}{Packet Gateway}
\newacronym{phy}{PHY}{Physical}
\newacronym{pbch}{PBCH}{Physical Broadcast Channel}
\newacronym[plural=\gls{mme}s,firstplural=Mobility Management Entities (MMEs)]{mme}{MME}{Mobility Management Entity}
\newacronym{prb}{PRB}{Physical Resource Block}
\newacronym{pss}{PSS}{Primary Synchronization Signal}
\newacronym{pucch}{PUCCH}{Physical Uplink Control Channel}
\newacronym{pusch}{PUSCH}{Physical Uplink Shared Channel}
\newacronym{embb}{eMBB}{Enhanced Mobile Broadband}
\newacronym{urllc}{URLLC}{Ultra Reliable Low Latency Communication}
\newacronym{fifo}{FIFO}{First In First Out}
\newacronym{rach}{RACH}{Random Access Channel}
\newacronym{ran}{RAN}{radio access network}
\newacronym{red}{RED}{Random Early Detection}
\newacronym{rlc}{RLC}{Radio Link Control}
\newacronym{est}{QUIC-EST}{QUIC-Enabled Scheduling and Transmission}
\newacronym{rlf}{RLF}{Radio Link Failure}
\newacronym{qoi}{QoI}{Quality of Information}
\newacronym{rrc}{RRC}{Radio Resource Control}
\newacronym{rrm}{RRM}{Radio Resource Management}
\newacronym{rr}{RR}{Round Robin}
\newacronym{rs}{RS}{Remote Server}
\newacronym{rsrp}{RSRP}{Reference Signal Received Power}
\newacronym{rss}{RSS}{Received Signal Strength}
\newacronym{rtt}{RTT}{round trip time}
\newacronym{rw}{RW}{Receive Window}
\newacronym{rx}{RX}{Receiver}
\newacronym{sa}{SA}{standalone}
\newacronym{sack}{SACK}{Selective Acknowledgment}
\newacronym{sap}{SAP}{Service Access Point}
\newacronym{sch}{SCH}{Secondary Cell Handover}
\newacronym{scoot}{SCOOT}{Split Cycle Offset Optimization Technique}
\newacronym{sdma}{SDMA}{Spatial Division Multiple Access}
\newacronym{sinr}{SINR}{Signal to Interference plus Noise Ratio}
\newacronym{sm}{SM}{Saturation Mode}
\newacronym{snr}{SNR}{Signal to Noise Ratio}
\newacronym{son}{SON}{Self-organizing network}
\newacronym{ss}{SS}{Synchronization Signal}
\newacronym{srs}{SRS}{Sounding Reference Signal}
\newacronym{sss}{SSS}{Secondary Synchronization Signal}
\newacronym{tb}{TB}{Transport Block}
\newacronym{tcp}{TCP}{Transmission Control Protocol}
\newacronym{tdd}{TDD}{Time Division Duplexing}
\newacronym{tdma}{TDMA}{Time Division Multiple Access}
\newacronym{tfl}{TfL}{Transport for London}
\newacronym{tm}{TM}{Transparent Mode}
\newacronym{prr}{PRR}{Packet Reception Ratio}
\newacronym{trp}{TRP}{Transmitter Receiver Pair}
\newacronym{tti}{TTI}{Transmission Time Interval}
\newacronym{ttt}{TTT}{Time-to-Trigger}
\newacronym{tx}{TX}{Transmitter}
\newacronym{ue}{UE}{User Equipment}
\newacronym{ul}{UL}{Uplink}
\newacronym{uml}{UML}{Unified Modeling Language}
\newacronym{um}{UM}{Unacknowledged Mode}
\newacronym{utc}{UTC}{Urban Traffic Control}
\newacronym{vm}{VM}{Virtual Machine}
\newacronym{rsrq}{RSRQ}{Reference Signal Received Quality}
\newacronym{rssi}{RSSI}{Received Signal Strength Indicator}
\newacronym{crs}{CRS}{Cell Reference Signal}
\newacronym{v2v}{V2V}{Vehicle-to-Vehicle}
\newacronym{v2i}{V2I}{Vehicle-to-Infrastructure}
\newacronym{v2n}{V2N}{Vehicle-to-Network}
\newacronym{v2x}{V2X}{Vehicle-to-Everything}
\newacronym{vn}{VN}{Vehicular Node}
\newacronym{dsrc}{DSRC}{Dedicated Short Range Communication}
\newacronym{ci}{CI}{context information}
\newacronym{qkp}{QKP}{Quadratic Knapsack Problem}
\newacronym{voi}{VoI}{Value of Information}
\newacronym{gps}{GPS}{Global Positioning System}
\newacronym{qos}{QoS}{Quality of Service}
\newacronym{qoe}{QoE}{Quality of Experience}
\newacronym{ml}{ML}{machine learning}
\newacronym{ahp}{AHP}{Analytic Hierarchy Process}
\newacronym{lidar}{LIDAR}{Light Detection and Ranging}
\newacronym{sumo}{SUMO}{Simulation of Urban MObility}
\newacronym{wave}{WAVE}{Wireless Access in Vehicular Environment}
\newacronym{c-its}{C-ITS}{Connected Intelligent Transportation System}
\newacronym{dash}{DASH}{Dynamic Adaptive Streaming over HTTP}
\newacronym{http}{HTTP}{HyperText Transfer Protocol}
\newacronym{nt}{NT}{non-terrestrial}
\newacronym{ntc}{NTC}{non-terrestrial communication}
\newacronym{ntn}{NTN}{non-terrestrial network}
\newacronym{haps}{HAPS}{high altitude platform station}
\newacronym{leo}{LEO}{Low Earth Orbit}
\newacronym{meo}{MEO}{Medium Earth Orbit}
\newacronym{geo}{GEO}{Geostationary Earth Orbit}
\newacronym{uav}{UAV}{unmanned aerial vehicle}
\newacronym{nsat}{nSAT}{Nanosatellite}
\newacronym{ehf}{EHF}{extremely high-frequency}
\newacronym{ioe}{IoE}{Internet of Everyone}
\newacronym{gan}{GaN}{Gallium Nitride}
 \newacronym{kpi}{KPI}{key performance indicator}
 \newacronym{mno}{MNO}{mobile network operator}
 \newacronym{plc}{PLC}{power line communication}
 \newacronym{isp}{ISP}{Internet service provider}
 \newacronym{sdr}{SDR}{software defined radio}
 \newacronym{iab}{IAB}{integrated access and backhaul}
 \newacronym{fso}{FSO}{free space optical}
 \newacronym{eon}{EON}{elastic optical network}
 \newacronym{ai}{AI}{artificial intelligence}
 \newacronym{noma}{NOMA}{non-orthogonal multiple access}
 \newacronym{irs}{IRS}{intelligent reflecting surface}
 \newacronym{oam}{OAM}{Operations, administration and management}
\newacronym{ws}{WS}{white space}
\newacronym{d2d}{D2D}{device-to-device}
\newacronym{owc}{OWC}{optical wireless communication}
\newacronym{vlc}{VLC}{visible light communication}
\newacronym{rf}{RF}{radio  frequency}
\newacronym{ti}{TI}{Tactile Internet}
\newacronym{vr}{VR}{virtual reality} \newacronym{ar}{AR}{augmented reality}

\usepackage{hyperref}

\begin{document}
\pagenumbering{gobble}


\title{QUIC-EST: A Transmission Scheme to Maximize VoI of Multi-Stream Correlated Data Flows}

\author{{Federico Chiariotti,~\IEEEmembership{Member, IEEE}, Anay Ajit Deshpande,~\IEEEmembership{Student Member, IEEE},\\ Marco Giordani,~\IEEEmembership{Member, IEEE},
Kostantinos Antonakoglou, Andrea Zanella,~\IEEEmembership{Senior Member, IEEE}, \\ Toktam Mahmoodi,~\IEEEmembership{Senior Member, IEEE}}

\thanks{
F. Chiariotti (corresponding author, email: fchi@es.aau.dk) is with the Department of Electronic Systems, Aalborg University, Denmark.
A. A. Deshpande, M. Giordani and A. Zanella (email: name.surname@dei.unipd.it) are with the Department of Information Engineering, University of Padova, Italy.
K. Antonakoglou  and T. Mahmoodi (email:name.surname@kcl.ac.uk) are with the Centre for Telecommunications Research, King's College London, United Kingdom. This project has received funding from the European Union's Horizon 2020 research and innovation programme under the Marie Skłodowska-Curie Grant agreement No. 813999 (WINDMILL).
}
}

\maketitle

\begin{abstract}
New advanced applications, such as autonomous driving and haptic communication, require to transmit multi-sensory data and require low latency and high reliability. These applications include. Existing implementations for such services have mostly relied on \emph{ad hoc} scheduling and send rate adaptation mechanisms, implemented directly by the application and running over UDP.
In this work, we propose a transmission scheme that relies on the features of the recently developed QUIC transport protocol, providing reliability where needed, and standardized congestion control, without compromising latency. Furthermore, we propose a scheduler for sensor data transmissions on the transport layer that can exploit the correlations over time and across sensors. This mechanism allows applications to maximize the \gls{voi} of the transmitted data, as we demonstrate through simulations in two realistic application scenarios.
\end{abstract}

\begin{picture}(0,0)(-33,-350)
\put(0,0){
\put(0,0){This paper has been submitted to IEEE for publication. Copyright may be transferred without notice.}}
\end{picture}


\glsreset{ti}
\glsreset{v2x}
\glsreset{voi}

\section{Introduction}
\label{sec:introduction}

The fifth generation (5G) of cellular networks is expected to provide extremely low latency, high reliability, and high throughput services, which appeal to interactive and demanding applications such as haptic communication, automated driving, and others~\cite{andrews2014will}. Along with the freedom of movement given by wireless connectivity, these applications generally require the timely and synchronized delivery of a multitude of sensor data and commands, to guarantee control effectiveness. 

For example, haptic communication allows users to interact with remote environments using haptic devices (i.e., haptic sensors and actuators) that exchange kinesthetic and tactile information. In the case of the closed-loop bilateral teleoperation systems, kinesthetic data is time-sensitive. Although stability control mechanisms can be employed to compensate for undesirable end-to-end delays that disturb the stability of such systems, this approach may deteriorate the \textit{transparency} of the service, i.e., the feeling of interactivity and, hence, the quality of telepresence \cite{lawrence1993}. A more transparent way to decrease the end-to-end delay, instead, consists in reducing the sensor data to be transmitted according to human perception model, but at the cost of a less accurate reconstructed signal at the receiver. \cite{Antonakoglou2018}.

Connected vehicles, on the other hand, will exchange data generated by on-board sensors via \gls{v2x} communications, to collaboratively build a richer context awareness and coordinate driving decisions~\cite{higuchi2019value}.
However, disseminating sensors' observations is expected to increase data traffic in vehicular networks by multiple orders of magnitude, thus potentially leading to congestion.

In order to operate effectively, these applications will need to implement the basic functions of the transport layer, i.e, retransmitting lost packets and limiting their send rate to avoid congestion. The  \gls{tcp} and \gls{udp} are the two traditional options, each with its pitfalls. 
By using \gls{tcp}, applications can delegate congestion control and retransmission to the transport layer, providing a simple and well-tested interface with standardized behavior. However, most congestion control mechanisms can create significant latency issues, and the requirement for in-order delivery can cause the head of line blocking problem, increasing the delay as received data is buffered waiting for the retransmission of earlier lost packets. In turn, \gls{udp} offers full flexibility to the applications, but leaves the burden of managing congestion and retransmissions to them.

In order to overcome the issues of both protocols, in this work we propose \gls{est}, a framework that combines the recently developed QUIC protocol with a suitable scheduling scheme. QUIC combines the ease of use and retransmission/control mechanisms of \gls{tcp} with \gls{udp}'s flexibility in the data delivery order, implementing different \emph{streams} that can be delivered in parallel and independently of each other and thus avoiding the head of line blocking issue and maintaining low latency. Implementers then have  freedom in scheduling the application data on the streams. We hence propose a multi-stream scheduling scheme that, leveraging the QUIC features, biases data transmissions as a function of the \gls{voi}~\cite{giordani2019investigating}. We define the \gls{voi} as a scalar value quantifying the utility of the data for the receiving application. The \gls{voi} takes into consideration the potential correlation in time and across different sensors, as well as the intrinsic value of their measurements. The performance of this proposal is tested in the haptic communication and \gls{v2x} scenarios, and we show that our approach guarantees better utility compared to traditional scheduler implementations.

The rest of the paper is organized as follows. Sec.~\ref{sec:quic} presents the QUIC protocol, the adaptations we introduce to the protocol, and Sec.~\ref{sec:sched} describes our proposed \gls{voi}-based scheduler. Sec.~\ref{sec:applications} presents two reference scenarios in which our system can be used, namely, autonomous driving and haptic communication, while we evaluate the performance of the \gls{voi}-based scheduler in Sec.~\ref{sec:performance_evaluation}. Finally, Sec.~\ref{sec:conclusions} concludes the paper.


\section{Adapting QUIC for Time-Sensitive Multi-Sensory Applications}\label{sec:quic}

The QUIC protocol~\cite{langley2017quic} was designed by Google to solve some of the latency issues that \gls{tcp} typically causes with Internet traffic. QUIC packets are encrypted by default and encapsulated in \gls{udp} datagrams, and the protocol is run directly on applications within the operating system kernel, making it easier to modify and~deploy. 

\begin{figure}[t]
    \centering
    \includegraphics[width=0.95\columnwidth]{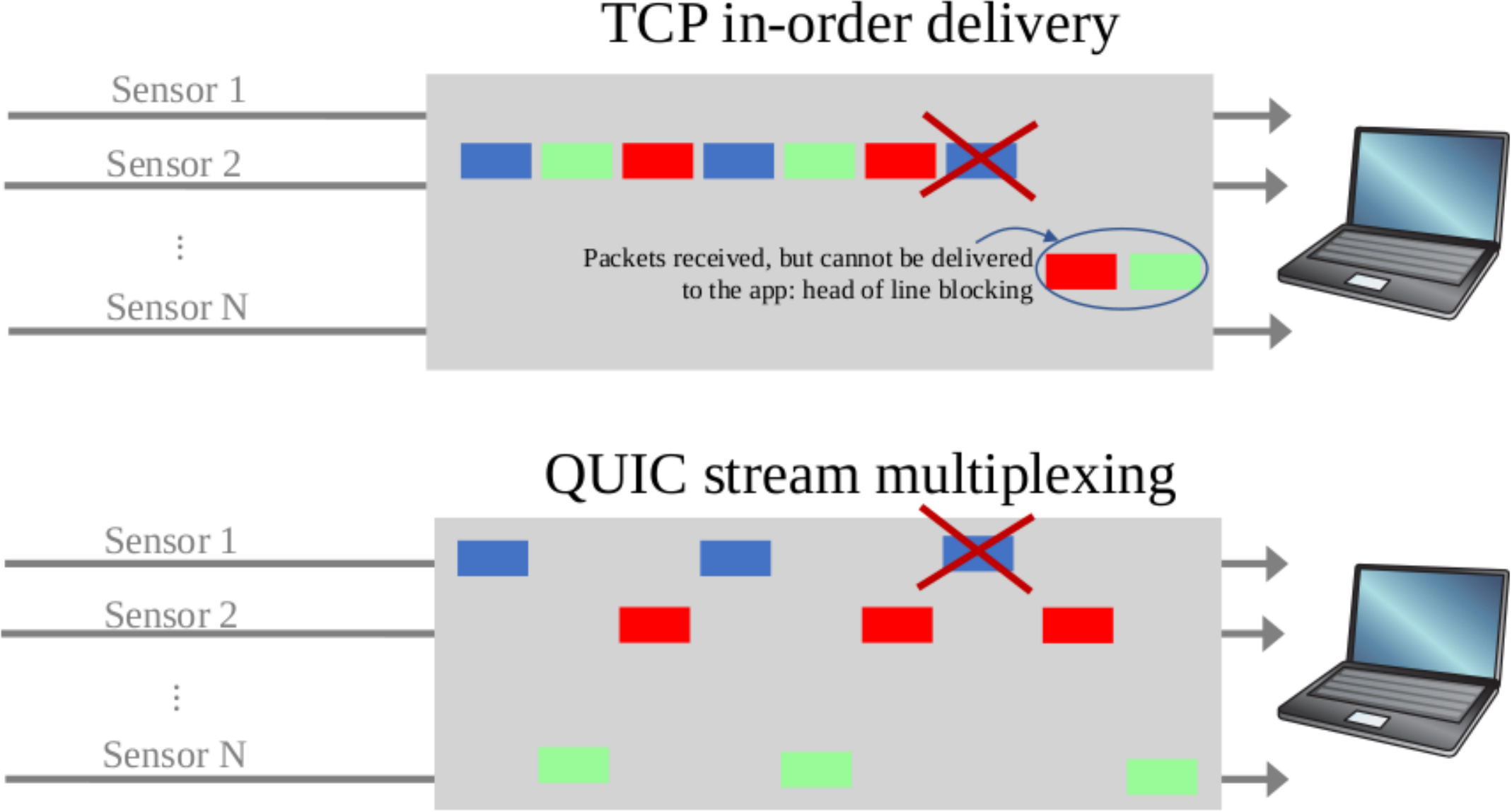}
    \caption{The head of line blocking problem and the stream-based solution.}
    \label{fig:hol}
\end{figure}

As well as providing no encryption, \gls{tcp} requires in-order delivery, interpreting transmitted data as a single stream and leaving the task of separating application-level objects to the application itself. In-order delivery can cause issues when multiple objects are transmitted over the same connection, such as for most Web pages. An error on one element of the page can indeed block other objects for a significant time, even though they were already received and could be displayed immediately. QUIC addresses the head of line blocking issue by adopting the same solution as the older \gls{sctp}, i.e., defining separate \emph{streams} of data within the same connection. Each stream is treated by the protocol as a logically separate data flow with in-order reliable delivery, independent of the other streams. Fig.~\ref{fig:hol} shows an example of the way QUIC handles multiple streams: while the loss of the blue packet also blocks the orange and green packets in \gls{tcp}, the logical separation between the streams allows QUIC to deliver the data.

QUIC was designed for Web traffic consisting of a potentially large number of logically independent objects to be delivered with the lowest possible latency. However, its features can be well-adapted to interactive multi-sensory applications in which sensing data needs to be broadcast, potentially with low delay, to preserve the user's \gls{qoe}. Unlike Web traffic, these applications do not usually require data from all available sensors, as they are built to be redundant. This makes the head of line blocking issue even more pressing, since the undelivered data might not even be necessary for successful operation.

Two examples of such applications are automated driving and haptic teleoperation, which typically use \gls{udp} at the transport layer, handling congestion and retransmissions at the application. 
In turn, we propose the \gls{est} scheme as a way of adapting QUIC to the multi-sensory application requirements.

In \gls{est}, each sensory reading can be considered as a separate object. As sensors produce several  readings per second, we propose to use not just a different stream for each sensor, but a \emph{different stream for each object}. Whenever all packets sent into a stream are acknowledged, the stream can be reused for a new object.
On the contrary, if a stream gets blocked by a packet loss and the data become stale, the send will transmit a \texttt{RESET\_STREAM} frame to tell the receiver to discard any existing out-of-order data received from that stream and start anew, as suggested in~\cite{shi2019dtp}.

\section{Value of Information-Based Scheduling}\label{sec:sched}

While the use of streams allows QUIC to transmit data from different sensors independently, the capacity of the connection might not be sufficient to deliver the data from all sensors within the required time. In this case, the choice of which sensory readings are transmitted and in which order becomes a central problem. The QUIC standard does not specify a scheduler for streams, leaving the possibility to implement a priority-based mechanism open. 

We then define a scheduling algorithm that aims at maximizing the \emph{effective \gls{voi}} at the receiver, while avoiding congestion in the connection. To this end, the algorithm needs to be fed with four types of information, namely: \emph{(i)} the (estimated) available capacity of the connection, \emph{(ii)} the size of the data with respect to the capacity, \emph{(iii)} the intrinsic \gls{voi} of the data, and \emph{(iv)} the correlation between the data generated by different sensors (which impacts the effective \gls{voi} of the transmitted data).

More specifically, we assume that, denoting by $N$ the number of objects generated in a batch by an application, the scheduler is given the following inputs.
\begin{itemize}
\item
The available capacity $C$ along the path. This information can be provided by the \gls{bbr} congestion control algorithm \cite{cardwell2016bbr}, which estimates the bottleneck capacity and minimum \gls{rtt} directly to avoid congestion, and promises a low transmission delay which is crucial for latency-sensitive applications. A possible alternative is the use of other latency-aware mechanisms such as the classic Vegas algorithm, estimating the capacity indirectly from the congestion windows. QUIC natively supports both mechanisms.

\item
The \emph{size vector} $\mathbf{s} \in \mathbb{N}^N$, which contains the sizes of the objects, in bits. Clearly, the amount of data sent over the connection should not exceed its capacity $C$, to avoid congestion.

\item  The \emph{weight vector} $\mathbf{v} \in [0; 1]^N$, which contains the \emph{intrinsic value} of the objects, i.e., the \gls{voi} when considering only that source. The intrinsic \gls{voi} is determined based on a number of factors, such as the position of the sensor  (e.g., front sensors in an autonomous vehicle are generally more informative than side sensors for driving decisions, or the finger sensors in a haptic application are more informative than wrist sensors), the current state of the process (e.g., the presence of an interesting  object in a camera's field of view, or the detection of a sudden gesture in haptic applications). The intrinsic \gls{voi} can also depend on the \emph{time correlation} of the sample process. If the process is slow varying, consecutive readings from the same sensor can be highly correlated and, hence, easily predictable by the receiver. Although the relation between the time since the last update from a sensor and the correlation with the new reading is highly application-dependent, it is often assumed to follow an exponential decrease~\cite{giordani2019investigating}. Some control applications have inbuilt compensation mechanisms for delay, which do not require new measurements until a certain time has passed, so the correlation for these cases can be modeled as a step function. A sigmoid function can then be used to model an imperfect compensation mechanism with a gentler degradation curve.
Given the specificities of the different applications, we assume that the  intrinsic \gls{voi} is determined by the application itself, and passed to the scheduling algorithm in the form of the weight vector. In the next section we will provide examples of how these values can be computed in two use-cases.

\item The \emph{cross-sensor correlation matrix} $\mathbf{W}\in [0; 1]^{N \times  N}$, which contains the correlation between objects. Indeed, if the application relies on multiple sensors, there is often a significant amount of redundancy in their information. For example, multiple cameras might have partially overlapping \glspl{fov}, or scalar sensors might be measuring correlated quantities. Therefore, the intrinsic \gls{voi} of some data may need to be discounted to account for the cross-sensor correlation, because the effective \gls{voi} of two correlated measurements can  be lower than the sum of the \glspl{voi} of the two measurements taken separately.  
\end{itemize}

\begin{figure}[!t]
 \centering
  \resizebox{0.99\columnwidth}{!}{
 \footnotesize{
\tikzset{  
block/.style    = {draw, rectangle, minimum height = 1em, minimum width = 1em}}
\begin{tikzpicture}[auto]
\coordinate(empty2) at (5, 0) {};
\node at (4.5,0) [draw, minimum width=4cm, minimum height=4cm,fill={gray!20}] (sched) {};
\node at (4.5,1) [draw, minimum width=3cm, minimum height=0.5cm,fill=white] (capacity) {Residual capacity};
\node at (4.5,-1) [draw, minimum width=3cm, minimum height=1.5cm,fill={blue!20}] (c1) {2};
\node at (4.5,-0.25) [draw, minimum width=3cm, minimum height=0.75cm,fill={red!20}] (c2) {3};
\node at (4.5,0.5) [draw, minimum width=3cm, minimum height=0.75cm,fill={red!20}] (c3) {1};
\node at (4.5,0.5) [draw,dashed, minimum width=5cm, minimum height=5.5cm] (quic) {};
\node at (6,2.75) [draw, minimum width=1cm, minimum height=0.5cm,fill={green!20}] (bbr) {BBR};
\node at (3,4) [draw, minimum width=1cm, minimum height=0.5cm,fill={green!20}] (app) {Appl.};
\node at (0,0) [draw, minimum width=1cm, minimum height=0.5cm,fill={green!20}] (s3) {Sensor 3};
\node at (0,0.75) [draw, minimum width=1cm, minimum height=0.5cm,fill={blue!20}] (s2) {Sensor 2};
\node at (0,-0.75) [draw, minimum width=1cm, minimum height=0.5cm,fill={red!20}] (s4) {Sensor 4};
\node at (0,1.5) [draw, minimum width=1cm, minimum height=0.5cm,fill={green!20}] (s1) {Sensor 1};
\node at (0,-1.5) [draw, minimum width=1cm, minimum height=0.5cm,fill={red!20}] (s5) {Sensor 5};
\node at (4.5,1.75) [minimum width=2cm, minimum height=1cm] (slab) {\textbf{Scheduling}};
\node at (4.5,3) [minimum width=2cm, minimum height=1cm] (qlab) {\textbf{QUIC socket}};
\node at (7.75,0) [draw, minimum width=1cm, minimum height=0.5cm,fill={blue!20}] (d1) {2};
\node at (8.5,0) [draw, minimum width=0.5cm, minimum height=0.5cm,fill={red!20}] (d2) {3};
\node at (9,0) [draw, minimum width=0.5cm, minimum height=0.5cm,fill={red!20}] (d3) {1};

\draw[->] (s1.east) to node[pos=0.4,above] {Data, VoI} (2.5,1.5);
\draw[-] (sched.east) to (d1.west);
\draw[->] (d3.east) to node[midway,above] {Sending}  (10.5,0);
\draw[->] (s2.east) to node[pos=0.4,above] {Data, VoI} (2.5,0.75);
\draw[->] (s3.east) to node[pos=0.4,above] {Data, VoI} (sched.west);
\draw[->] (s4.east) to node[pos=0.4,above] {Data, VoI} (2.5,-0.75);
\draw[->] (s5.east) to node[pos=0.4,above] {Data, VoI} (2.5,-1.5);
\draw[->] (bbr.south) to node[midway] {Capacity} (6,2);
\draw[->] (app.south) to node[pos=0.1] {Sensor correlation} (3,2);
\end{tikzpicture}
}
}
\caption{Basic components of the framework and main data exchanges. In the figure, the data from sensors 1 and 5 is discarded, while the data from sensors 2, 3, and 4 is sent in that order.}
\label{fig:scheduler}
\end{figure}
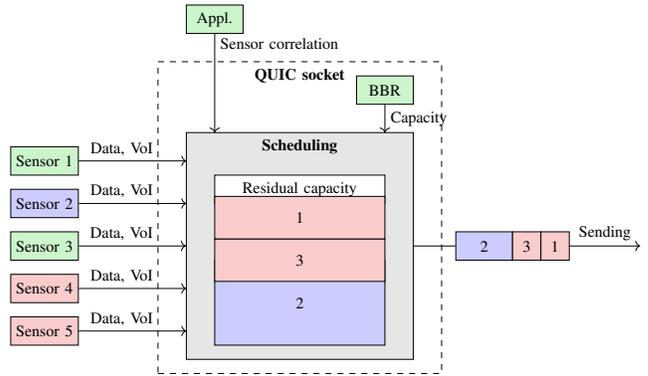

Finding the optimal schedule can then be reduced to finding the set of objects that maximizes the \gls{voi} while respecting the delay requirements. If we limit the analysis to couples of objects, i.e., we do not consider the effects of triplets of correlated objects, this is an instance of the \gls{qkp}~\cite{pisinger2007quadratic}, which is NP-hard, but efficient heuristics to solve it have been developed.  Fig.~\ref{fig:scheduler} shows the basic structure of the proposed scheduling framework: multiple sensors write data with a given \gls{voi} to a QUIC socket, and the application supplies the cross-sensor correlation matrix $\mathbf{W}$. The available capacity is read from the BRR estimate, and the scheduler finds the optimal set of objects that can be delivered before the next sensor update, sending them through the connection as fast as congestion control allows. If the connection is lossy or time-varying, scheduling decisions can be revised based on what was already sent and recomputing the solution to the problem.

To the best of our knowledge, the scheduling of multi-sensory data on the transport layer is a new research topic, which requires the study of the application and sensor features as well as the dynamics of the end-to-end capacity. \gls{est} is a first step in that direction, considering correlated measurements in time and across multiple sensors and using congestion control to estimate the capacity. The scheduling framework is relatively simple, but it can support a wide range of applications, guaranteeing reliability and maximizing the delivered \gls{voi}.

\section{Application Scenarios for QUIC-EST}
\label{sec:applications}

The scheduling problem is a well-studied subject for multi-stream applications~\cite{bai2010uplink} at the medium access layer, but considering temporal and cross-sensor correlation is a relatively new idea, which has never been applied at the transport layer, to the best of our knowledge.
In the following, we give two examples of applications that might benefit from the proposed QUIC-EST modifications and VoI-based scheduler, namely autonomous driving (Sec.~\ref{sub:quic_for_v2x}) and  haptic communication (Sec.~\ref{sub:quic_for_ti}).

\subsection{QUIC-EST for Autonomous Driving} 
\label{sub:quic_for_v2x}

\begin{figure*}[t]
    \centering
    \includegraphics[width=0.99\textwidth]{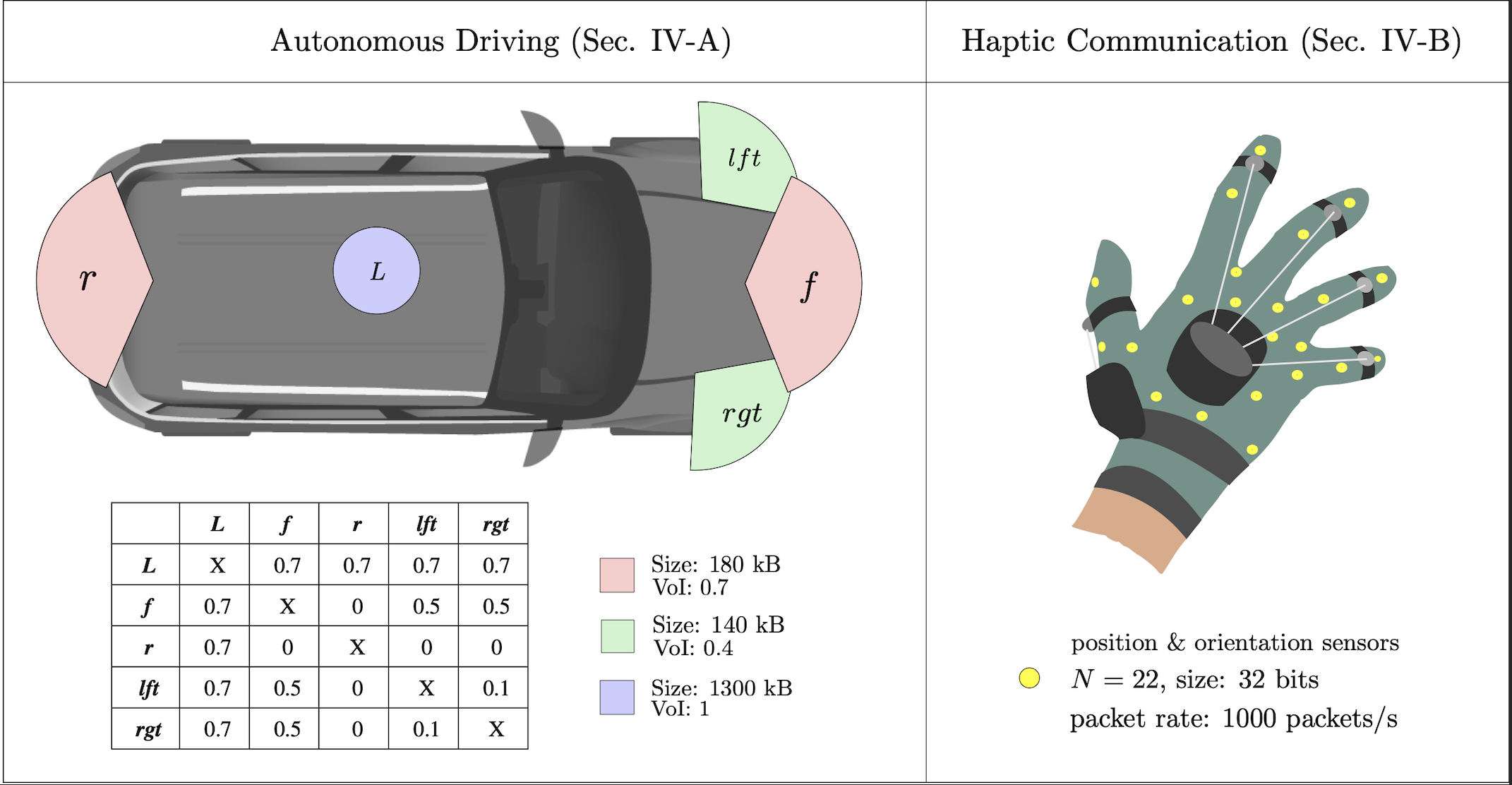}
    \caption{Scheduling input parameters for the autonomous driving (left) and haptic communication (right) scenarios.}
    \label{fig:scenarios}
\end{figure*}


The autonomous driving scheduling problem can be reformulated as follows.
\smallskip

\emph{Size vector.}
First, we define the size vector, which depends on the type of automotive sensor that is considered, the rate at which observations are captured, and their resolution. For example, while for low resolution \gls{lidar} sensors the data rate is relatively low at around 50 kbps, for high resolution \gls{lidar} sensors, such as the Velodyne HDL-64E, the required data
rate is around 30 Mbps. 
For camera images, the data rate ranges from around 10 Mbps to  500 Mbps depending on the resolution~\cite{va2016millimeter}, even though JPG compression can reduce the image size by several orders of magnitude.

In this work, we consider $N=5$ sensors: one camera on the vehicles' top left corner ($lft$), one  on the top right corner ($rgt$), one  on the front size ($f$), one  on the rear side ($r$), and one \gls{lidar} on the rooftop of the car ($L$).
The size of sensors' observations are calculated based on the nuScenes dataset~\cite{caesar2020nuscenes}, which carries a full autonomous vehicle sensor suite, assuming a 1 Byte pixel encoding and  JPG compression for the camera images: we consider a size for the front/rear cameras of 180 KB, for the lateral cameras of  140 KB, and for the \gls{lidar} of 1300 KB, as depicted in Fig.~\ref{fig:scenarios} (left).
\smallskip

\emph{Intrinsic \gls{voi}.}
In our scheduler, data transmission is discriminated not only based on the intrinsic characteristics of the different sensors  but also on the correlation among consecutive observations.
This information is contained in the weight vector. In particular, we reasonably expect that the \gls{lidar} would be  more valuable compared to automotive cameras because it can provide a three-dimensional (rather than bi-dimensional) representation of the environment, and can work efficiently in different weather/time conditions.
Also, we assume that different cameras might have different priorities depending on the characteristics of the environment in which the vehicles move (e.g., for the majority of the time, lateral cameras in the highway scenario likely make background observations with very similar characteristics, while frontal/rear cameras' acquisitions might incorporate more valuable information). Based on this assumption, the correlation vector $\mathbf{v}\in[0,1]^N$ can be built empirically, resulting in the values shown in Fig.~\ref{fig:scenarios} (left).

The \gls{voi} should also depend on  the temporal obsolescence of the information.
Based on the results in~\cite{giordani2019investigating}, we suggest to use an exponential function that depends on the relative age of information, i.e., the time between the generation and reception of the information, and a temporal decay parameter that is proportional to the  delay sensitivity of the observation, i.e., the temporal horizon over which that piece of information is considered valuable.
\smallskip

\emph{Cross-sensor correlation.}
As the name suggests, the cross-sensor correlation matrix represents the degree of correlation among the different sensors. For cameras, the correlation depends on the \gls{fov}. We assume that the rear camera is uncorrelated to the rest of the cameras, very little overlap between lateral cameras, and partial overlap between lateral and front cameras. 
On the other hand, \gls{lidar} sensors operate through 360-degree rotations and their observations can be highly correlated/redundant with those of the cameras. 
The correlation matrix is hence structured  as displayed in Fig.~\ref{fig:scenarios} (left).


\subsection{QUIC-EST for Haptic Communication} 
\label{sub:quic_for_ti}


For the haptic communication scenario, the scheduling problem can be reformulated as follows.
\smallskip

\emph{Size vector.} In this scenario, the size vector should depend on the amount of sensors and actuators integrated on the haptic devices. The CyberGrasp~\cite{CyberGrasp2013} device  combines a haptic glove that is sensing orientation and movement of the hand using 22 sensors and an exoskeleton with five kinesthetic actuators for providing force feedback to the user. Taking into account only the 22 movement sensors that transmit one floating-point value each (i.e., typically 32 bits using the IEEE 754 standard) with the typical 1 kHz sampling rate, for two haptic devices (one for each hand) we have $N = 44$, resulting in a 1.4 Mbps data stream, as represented in Fig.~\ref{fig:scenarios} (right).
\smallskip

\emph{Intrinsic \gls{voi}.} In order to determine the \gls{voi} of each haptic data sample generated by the haptic device's sensors, we rely on the psychophysical aspects of human perception. More specifically, we can use Weber's law of \gls{jnd}, as in the deadband transmission algorithm in \cite{Hinterseer2005}, which can be applied in position, velocity and force data. The \gls{voi} is then given by the difference between the last transmitted sample from that sensor and the current value, which can be easily computed by the sending application and given to the scheduler; sensors have the same inherent \gls{voi}, but the actual value of the information depends on how novel it is with respect to the one currently available to the receiver. This definition implicitly includes the time correlation between samples, as the difference between consecutive samples will usually be small, but then grow with time consistently with the age of the data. 
\smallskip


\emph{Cross-sensor correlation.} In the haptic communication case, the flexibility of a robotic hand makes the relation between different sensors strongly dependent on their position. If the hand is grasping an object, the correlation between sensors will be different from when it is at rest. Consequently, we cannot give a constant cross-sensor correlation matrix based on the sensors' positions, like we did in the vehicular case; if the application can compute the instantaneous correlation between sensors in real time, the scheduler will use it to improve performance, but if it cannot, it will simply consider the measurements independent.

\section{Performance Evaluation} 
\label{sec:performance_evaluation}

In this section, we present a performance evaluation of QUIC-EST, comparing our scheduling to different algorithms and showing its benefits in terms of delivered \gls{voi}. We evaluate the system in the two scenarios presented in Sec.~\ref{sec:applications}, with extremely different features. While realistic, the assumptions about the two scenarios are arbitrary, and the purpose of showing them is to illustrate the methodology from a qualitative perspective, rather than giving a complete quantitative assessment of the scheme. The autonomous vehicle in the first scenario transmits only 10 frames per second but with a maximum rate of 155.2 Mbps; on the other hand,  haptic communication has a maximum rate of just 1.4 Mbps, but its sample frequency is 100 times faster, i.e., 1 kHz. Furthermore, while the haptic communication scenario has 44 different sensors that need to be scheduled, the autonomous driving scenario only has 5. 
We can consider these scenarios as examples of the two types of traffic patterns supported by 5G: the high-throughput \gls{v2x} traffic, which can tolerate a higher latency due to the lower frame rate, fits the requirements of \gls{embb} traffic, while the low-throughput haptic communication traffic, which requires millisecond latency, is a perfect example of \gls{urllc}.

In both scenarios, we study the average \gls{voi} as a function of the available (constant) capacity. 
We compare four different scheduler implementations:
\begin{itemize}
 \item \emph{\gls{fifo}.} This is the default QUIC scheduler, which transmits pieces of data in the same order they were received from the application. It limits transmission to the achievable send rate, discarding any objects that would exceed the connection's capacity. We consider this as a baseline, as its behavior is similar to \gls{tcp}, without the head of line blocking.
 \item \emph{\gls{voi}-based.} This scheduler considers the \gls{voi} as the decision factor for transmitting objects that fit the transmission capacity. It is an instance of the classic knapsack problem, as it does not consider cross-sensor correlation or even temporal correlation between values, but only the intrinsic value of each sensor.
 \item \emph{Cross-sensor \gls{voi}.} This scheduler considers cross-sensor correlation, but neglects the temporal correlation. It is equivalent to the optimal scheduler if subsequent measurements from the same sensor are  independent, and to the \gls{voi}-based scheduler if the sensors' measurements are independent as well.
 \item \emph{\gls{est}.} The scheduler considers \gls{voi} as well as time and cross-sensor correlation. It is equivalent to the \acrlong{qkp}, and gives the best performance if we consider the full application model.
\end{itemize}

\begin{figure}[!t]
 \centering
 \includegraphics{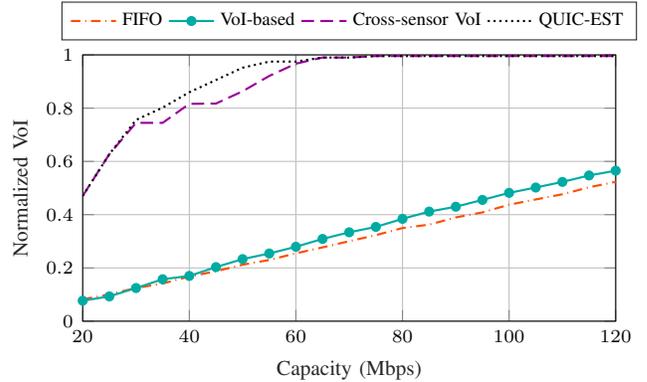}
 \caption{Comparison between schedulers in the autonomous driving scenario in terms of the normalized \gls{voi} as a function of capacity.}
 \label{fig:v2x_one}
\end{figure}

\begin{figure}[!t]
 \centering
 \includegraphics{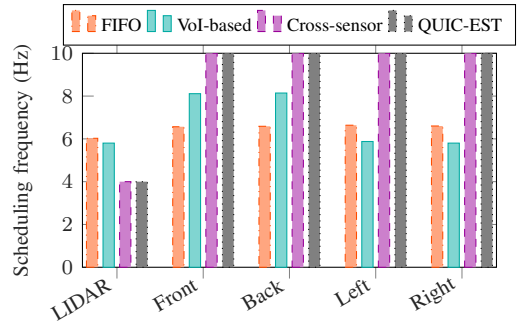}
 \caption{Average update frequency for the different schedulers in the autonomous driving scenario, with $C=100$ Mbps.}
 \label{fig:v2x_two}
\end{figure}

The performance for the autonomous driving scenario is shown in Fig.~\ref{fig:v2x_one}: as we consider capacities from 20 to 120 Mbps, the normalized \gls{voi}, defined as the average \gls{voi} divided by the \gls{voi} for a connection with infinite capacity, grows for all schedulers, but cross-sensor correlation is clearly the most important factor. As Fig.~\ref{fig:v2x_two} shows, this stark difference is due to the frequency at which the schedulers pick LIDAR frames, which are large and highly correlated with data from the cameras, as LIDAR has a $360$-degree \gls{fov}. If capacity is limited, prioritizing LIDAR frames leads to sending fewer camera frames, but while camera frames have a high joint value, LIDAR provides highly correlated data with relatively little new information. Schedulers that consider cross-correlation only send a limited number of LIDAR frames, maximizing the joint \gls{voi} and making the best use of the capacity.

For the haptic communication scenario, we consider the \gls{voi} as a logistic function of the difference between the current sample and the last transmitted one. We used realistic haptic traffic model parameters from~\cite{abu2009empirical} and cautiously selected a \acrfull{jnd} value of 5\% of the dynamic range of the sensors. Accordingly, we simulate each sensor as an independent Gauss-Markov process, setting $\sigma=2.15$\% of the dynamic range to fit the empirical model from the paper. The \gls{voi} is then given by a logistic function with a center $x_0=1.65\sigma$ and a sharpness $k=10$. We chose these values to ensure that all sensors with differences over the \gls{jnd} are prioritized, while sensors that are close to the threshold can be sent if the capacity allows it. In the haptic communication scenario, we have no cross-sensor correlation, and all sensors have the same intrinsic \gls{voi}, so the \gls{fifo}, \gls{voi}-based, and cross-sensor \gls{voi} schedulers are equivalent.

\begin{figure}[!t]
 \centering
 \includegraphics{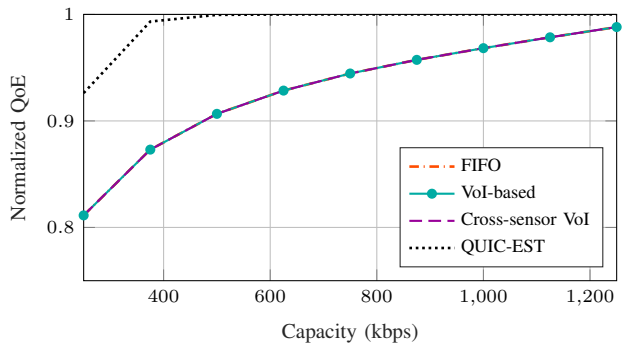}
 \caption{Comparison between schedulers in the haptic communication scenario in terms of the normalized \gls{qoe} as a function of capacity.}
 \label{fig:ti_perf}
\end{figure}

As Fig.~\ref{fig:ti_perf} shows, the \gls{est} scheme can achieve an almost perfect \gls{qoe} (i.e., most of the sensors are under the \gls{jnd} threshold most of the time) even at less than a third of the capacity needed to send all packets. In this case, the time correlation is critical: all other schedulers, which do not use a \gls{jnd}-based value, have a lower performance. This gain is the minimum for the scheme, as there are no cross-sensor correlations to exploit, and the introduction of more complex models that can take them into account could further decrease the resources that the application would need to request.

\section{Conclusions} 
\label{sec:conclusions}

In this paper, we have presented \gls{est}, a flexible transport-layer protocol based on the emerging QUIC protocol and meant for multi-sensory applications with time-sensitive data. We showed that this scheme, combined with a new scheduling mechanism that biases transmission decisions based on VoI and temporal/cross-sensor correlation, can be adapted to widely different applications with good results, using autonomous driving and haptic communication as our Future Internet use cases. At the moment, the transport layer is still an unexplored topic for this kind of applications, as research has focused on enabling them with single wireless links, but it is a crucial factor as their deployment in real networks begins. 

There are several possible avenues of future work which we plan to pursue. First, we will test \gls{est} in a more realistic setting, using data from real applications and full system-level simulation to give it a more robust evaluation. Another key aspect is reliability, as the distribution of capacity might play a role in scheduling time-varying scenarios, such as most wireless networks. Finally, the combination of this protocol with network slicing techniques would provide application with reliability guarantees, which can be fundamental for safety-critical applications such as \gls{v2x} for autonomous~vehicles.

\bibliographystyle{IEEEtran}
\bibliography{bibliography.bib}

\begin{IEEEbiographynophoto}{Federico Chiariotti} [SM'15,M'19] received his Ph.D. in Information Engineering in 2019 from the University of Padova, Italy. He is currently a postdoctoral researcher at Aalborg University, Denmark, after working as a research intern at Nokia Bell Labs in Dublin, Ireland. His work received the Best Paper Award in 4 conferences, including the 2020 IEEE INFOCOM WCNEE workshop. His research focuses on the latency-oriented design of networking protocols.\end{IEEEbiographynophoto}%
\begin{IEEEbiographynophoto}{Anay Ajit Deshpande} [SM'19] received his Bachelors in Electronics and Communication Engineering in 2016 from VIT University, India and Masters in Information and Communication Engineering in 2019 from Technische Universität Darmstadt, Germany. He is currently a PhD student and a Marie Curie Fellow under H2020-MSCA-ITN-WindMill at University of Padova, Italy. His research focusses on anticipatory techniques in wireless networks. \end{IEEEbiographynophoto}%
\begin{IEEEbiographynophoto}{Marco Giordani} 
[M'20] received his Ph.D. in Information Engineering in 2020 from the University of Padova, Italy,  where he is now a postdoctoral researcher and adjunct professor.
He visited  NYU and TOYOTA Infotechnology Center, Inc., USA.
In 2018 he received the “Daniel E. Noble Fellowship Award” from the IEEE Vehicular Technology Society. His research focuses on protocol design for 5G/6G mmWave cellular and vehicular networks.
\end{IEEEbiographynophoto}%
\begin{IEEEbiographynophoto}{Konstantinos Antonakoglou} received the M.Sc. degree in Control and Computing from the National and Kapodistrian University of Athens in 2014 and the Ph.D. degree in Telecommunications from King’s College London in 2020. He is currently a Research Associate at King’s College London and his research interests include haptic communication systems over networks, mobile cloud computing and edge-assisted AI systems.
\end{IEEEbiographynophoto}%
\begin{IEEEbiographynophoto}{Andrea Zanella} [S'98-M'01-SM'13] is Full Professor at the Department of Information Engineering (DEI), University of Padova (ITALY). He received the Laurea degree in Computer Engineering in 1998 from the same University and the PhD in 2001. During 2000, he was visiting scholar at the University of California, Los Angeles (UCLA), with Prof. Mario Gerla's research team. Andrea Zanella is one of the coordinators of the SIGnals and NETworking (SIGNET) research lab. His long-established research activities are in the fields of protocol design, optimization, and performance evaluation of wired and wireless networks. 
.\end{IEEEbiographynophoto}%
\begin{IEEEbiographynophoto}{Toktam Mahmoodi} received the B.Sc. degree in electrical engineering from the Sharif
University of Technology, Iran, in 2002, and the Ph.D. degree in telecommunications from
King’s College London, U.K, in 2009. Toktam is
currently Head of the Centre for Telecommunications Research at the Department of
Engineering, King’s College London. Her research interests include mobile communications,
network intelligence, and low latency networking.\end{IEEEbiographynophoto}%

\end{document}